\documentclass[aps,pre,twocolumn,superscriptaddress,groupedaddress]{revtex4}

\usepackage[utf8]{inputenc}
\usepackage[english]{babel}

\usepackage{hhline}
\usepackage{indentfirst}
\usepackage{fancyhdr}
\usepackage[final]{showkeys}
\usepackage{graphicx}
\usepackage{subfig}
\usepackage{amssymb}
\usepackage{comment}
\usepackage{latexsym}
\usepackage{amsthm}
\usepackage{amsmath}
\usepackage{eucal}
\usepackage{amsfonts}
\usepackage{mathtools}
\usepackage{esvect}
\usepackage{pgf,tikz}
\usepackage{mathrsfs}
\usepackage{epstopdf}
\usetikzlibrary{arrows}
\usepackage{enumerate}
\usepackage{epigraph}
\usepackage[a4paper, left=1.5cm, right=1.5cm, top=2cm, bottom=2cm]{geometry}
\usepackage{floatpag}
\usepackage{comment}

\citeindextrue

\newcommand{\braket}[1]{\ensuremath{\left\langle #1 \right\rangle}}

\begin{document}

\title{%Geometric approach of  critical
Magnetization profiles at the upper critical dimension \\ as solutions of the integer Yamabe problem}

\author{Alessandro Galvani}
\affiliation{SISSA and INFN, Sezione di Trieste, Via Bonomea 265, I-34136 Trieste, Italy}
%\affiliation{INFN, Sezione di Trieste, I-34151 Trieste, Italy}
%\date{}
%\DeclareUnicodeCharacter{221E}{∞}

\author{Giacomo Gori}
\affiliation{Institut f\"ur Theoretische  Physik,  Universit\"at Heidelberg, D-69120  Heidelberg,  Germany}
\affiliation{CNR-IOM DEMOCRITOS Simulation Center and SISSA, Via Bonomea 265, I-34136 Trieste, Italy}

\author{Andrea Trombettoni}
\affiliation{Department of Physics, University of Trieste, Strada Costiera 11, I-34151 Trieste, Italy}
\affiliation{SISSA  and  INFN,  Sezione  di  Trieste, Via Bonomea 265, I-34136 Trieste, Italy}
\affiliation{CNR-IOM DEMOCRITOS Simulation Center and SISSA, Via Bonomea 265, I-34136 Trieste, Italy}

\begin{abstract}
We study the connection between the magnetization profiles of models described by a scalar field with marginal interaction term in a bounded domain and the solutions of the so-called Yamabe problem in the same domain, which amounts to finding a metric having constant curvature.
Taking the slab as a reference domain, we first study the magnetization profiles at the upper critical dimensions $d=3$, $4$, $6$ for different
(scale invariant) boundary conditions.
By studying the saddle-point equations for the magnetization, we find general formulas in terms of Weierstrass elliptic functions, extending exact results known 
in literature and finding new ones for the case of percolation. The zeros and poles of the Weierstrass elliptic solutions can be put in direct connection with the boundary conditions. We then show that, for any dimension $d$, the magnetization profiles are solution of the corresponding integer Yamabe equation at the same $d$ and with the same boundary conditions.
The magnetization profiles in the specific case of the $4$-dimensional Ising model with fixed boundary conditions are compared with Monte Carlo simulations, finding good agreement. These results explicitly confirm at the upper critical dimension recent results presented in \cite{gori}. 
%This shows that the saddle-point equation for the critical magnetization at the upper critical dimension is equivalent to a purely geometric description, where the euclidean metric is conformally altered to obtain a space with uniform negative curvature. 
%There, general correlation functions of the magnetization field in a bounded domain have been conjectured to be related to the solution of a fractional Yamabe problem, which with vanishing anomalous dimension becomes the integer problem here considered. 
\end{abstract}

\maketitle
\section{Introduction}

Within the field of critical phenomena, 
%there are many reasons to focus one's attention
a research line that attracted considerable attention along the years is %provided by
the study of critical systems in presence of boundaries %, i.e. of of boundary critical phenomena 
\cite{Diehl1997,Domb2000}. Among the motivations for such a study, the first 
%first of all, it is necessary to compare 
is to correctly compare theoretical predictions with experimental results or %interpret 
numerical simulations %, unavoidably 
coming from finite systems. In general, the introduction of a boundary 
%can shift the critical temperature and 
 alters thermodynamic properties and sub-leading  corrections to bulk values have been extensively studied \cite{%Chamati1996,
Brankov2000}. Techniques such as finite-size scaling can be used to extract critical exponents and other important quantities \cite{Cardy1996}. In two-dimensional systems, the coefficient of the next-to-next-to-leading order in the free energy is related to the central charge of the theory \cite{cardy88} and the study of boundary conformal field theory developed in the last decades as a very active field of research \cite{CardyBCFT}.

%The field is %also quite rich
%indeed much more than the study of approximation %to the properties of homogeneous systems: 
A remarkable property emerging in critical bounded domains is that a given bulk universality class corresponds to several surface universality classes, making boundary critical phenomena a rich and well studied playground.
In many cases, boundaries %are not just a nuisance, but 
are the source of very interesting effects 
and phenomena. %the phenomenon being considered. The most well-known case 
A major example is the thermodynamic Casimir effect: it has been studied thoroughly to obtain universal scaling function and Casimir amplitudes \cite{Vasilyev2009}. The interest in the Casimir effect has shifted the attention from semi-infinite systems \cite{Bray1977,Diehl1980,Diehl1981,Lubensky1975} to the geometry of a slab \cite{Grueneberg2008,Diehl2017,Gambassi2006a,Gambassi2006,Vasilyev2007}. Obtaining the critical magnetization profile \cite{Krech1997,Vassilev2018} is 
%an important piece of this puzzle.
a crucial piece of information in this line of research.

%our motivations: 1) trovare formule valide per ogni d e per varie BC; 2) verificare Gori

The study of critical phenomena with a field theoretical approach \cite{ZinnJustin}
naturally highlights the role of the dimensionality which rules the
relevance of fluctuations at the critical point.
For $d$ matching the so-called upper critical dimension $d_c$ \cite{Cardy1996,Mussardo}
the critical theory becomes typically tractable, %but still interesting, 
yielding 
in a controlled way the mean-field approximation 
and the typical logarithmic corrections 
on top of it \cite{LeBellac,Amit}. If critical exponents at the upper critical dimension are found to be just the mean-field ones, in bounded systems there is the possibility to study inhomogeneous, space-dependent quantities and corrections due to the presence of boundaries.
%at the mean-field approximation can provide
%even more insights, for the appearance of %non-trivial inhomogeneities
%is essential part of experimental and theoretical investigations.

Here we are going to focus on magnetization profiles of %$O(N)$ 
 models at their upper critical dimension. 
%Besides being exact above the upper critical dimensions, 
They can be a useful starting point for calculation of critical magnetization profiles at lower dimensions, as they give the background around which (non-logarithmic) fluctuations can be added. Moreover, an advantage of being at the upper critical dimension is that one can obtain, by a saddle-point treatment, analytical expressions 
to be compared with numerical results. 

Our goal in this paper is threefold. First, 
we obtain magnetization profiles in a unified framework for different boundary conditions, by considering the analytic structure of the saddle-point solutions. 
This will allow us to recast known results in a more accessible and compact way
while deriving novel predictions, e.g., for percolation at its upper critical dimension. For the Ising model in four dimensions we test the findings obtained by the discussed approach by comparing them with results of Monte Carlo simulations.

Another main objective is to provide
a geometric interpretation of the results for the critical magnetization profiles at the upper critical dimension. The main idea is to write the field theory equations for the magnetization profiles as a variational solution of a geometric problem.  
In this way, {our aim} of finding the critical profiles 
is explicitly 
put in connection with the solution of a
celebrated problem in differential geometry:  
the Yamabe problem \cite{Yamabe1960}, 
which amounts to 
finding a metric, in the same conformal class of another given metric, that makes the scalar curvature constant.
%The next step %we take is to embed this field-theory-borne variational 
%approach into a bigger picture. 
The Yamabe problem in its various generalizations
has been the object of intense mathematical 
work in the last decades \cite{Lee1987, MarGonzalez2010}. It is also related to 
general relativity, when one looks for 
solutions of Einstein field equations which are conformally flat or more generally within a conformal class \cite{AKUTAGAWA}. 
A discussion of the connection between the (mean-field) Landau-Ginzburg equations and the Hilbert-Einstein action functional for pure gravity in presence of cosmological term is presented in \cite{KHOLODENKO}.

As a third motivation, the present study intends to explicitly verify the validity at the upper critical dimension of results presented in \cite{gori}. There, a hypothesis---referred to as ``uniformization''---is put forward to relate the critical magnetization profile of a bounded domain with a metric factor $\gamma(\mathbf{x})$, which is the solution of a fractional Yamabe equation in the same domain. This metric is used to compute distances between points, which are then used to construct spin-spin correlation functions. % at different spatial points from 
%the effects of boundaries at the critical point can be 
%accounted for by a geometric description, which in the current 
%work will be pursued at $d=d_c$ yielding agreement.
In this ``critical geometry'' approach, the %less correlated
points close to the boundary are put
at larger distances by inflating locally
and isotropically the original metric. In \cite{gori} the uniformization hypothesis was supported by numerical simulations of the three-dimensional
Ising model, given the lack of analytical results for this model. It is important to notice that below the upper critical dimension a nonvanishing anomalous dimensions $\eta$ arises: in order to take this into account, the Yamabe equation has to be modified, by introducing a fractional power of the Laplacian operator, obtaining the fractional Yamabe equation \cite{MarGonzalez2013}. But, at the upper critical dimension $\eta=0$ 
and the fractional Yamabe equation reduces to the standard Yamabe equation, which for clarity we call the {\it integer} Yamabe equation, since in it only the usual Laplacian enters. Therefore the study of critical magnetization profiles at $d=d_c$ allows us to test, analytically and numerically, the validity of the ``critical geometry'' approach in a solvable case.

We observe that the relation between  critical 
magnetization profiles and the integer Yamabe equation holds in any bounded domain. We specialize to the slab geometry since it is convenient for both analytical calculations and  simulations and for its relevance for the Casimir effect. We will comment on the general case when useful for our presentation.

%Noticing that the same complex function can provide various boundary conditions, we hypothesize that the same could be done to extend the results of \cite{gori} to different boundary conditions making the extent of our geometric approach much broader
%and fully integrated with the current body of knowledge.

%Then, we are going to summarize the description of bounded critical phenomena using the Yamabe equation, as introduced in \cite{gori}, in order to see how it reproduces the mean-field results.

%Here, we are going to show how some previous results on mean field magnetization profiles and boundary conformal field theory can be recovered and expanded upon with the use of the critical geometry description introduced in \cite{gori}. We will show that the mean field equation satisfied by the order parameter coincides with the Yamabe equation at the upper critical dimension.
%The effect of a boundary is greatly enhanced at the critical point, where the infinite correlation length causes even regions deep in the bulk to be affected by it.

The plan of the paper is the following: first we will consider Landau-Ginzburg actions with { marginal interaction, for various dimensions:} in each case, we will obtain the saddle-point equation at the corresponding upper critical dimension; solutions will be given for the possible boundary conditions (BC). These equations will then be generalized through the introduction of the Yamabe equation. After solving it for a slab in arbitrary dimension, we recover the previous solutions. The theoretical magnetization profiles are then compared to the numerical solution of the Ising mean field equation, and to the results of a Monte Carlo simulation of the four-dimensional Ising model.

\section{Magnetization profiles}\label{magprofiles}

In the following, %attention will be devoted to the 
we study exact solution for the critical magnetization 
profiles for models living at
their upper critical dimension.
{We write the action of a scalar field with marginal coupling and $\mathbb{Z}_2$ symmetry as}%at the upper critical dimension, for an order parameter with $\mathbb{Z}_2$ } symmetry has the 
\begin{equation}
 S=\int d^dx\left[ \frac 1 2\phi(-\bigtriangleup)\phi +\frac 1 2 \mu^2\phi^2 + 
 g\, c_d\, (\phi^2)^{\frac{d}{d-2}} \right],\label{action}\end{equation}
 where the field $\phi=\phi(\mathbf{x})$ with $\mathbf{x}\in\Omega$ and $\Omega\subset\mathbb{R}^d$ is a bounded domain. {As usual, the operator $-\bigtriangleup$ is the positive definite Laplacian. Notice that the exponent of the potential term is integer for $d=3,4$, which are the dimensions we are going to focus on. The case of percolation, considered later, has to be treated separately with the introduction of a similar action in $d=6$. The action can be generalized to a vector field $\vec{\phi}$ with $N$ components: in this case, $\phi^2$ in the potential term is replaced by $\sum_{i=1}^N \phi_i^2$. } {The factor 
in the interaction constant } 
\begin{equation}
c_d \equiv \frac{(d-2)^2}{8}    
\end{equation}%ruling interaction 
has been chosen for convenience, 
as it will appear later. Since in $d$ dimensions the field has scaling dimension $\Delta_{\phi}=\frac{d-2}{2}$, the {interaction term in} 
\eqref{action} is written so that the action is at the  upper critical dimension. The mass term 
$\propto \phi^2$ vanishing at the critical 
point is also included~\cite{Cardy1996,Mussardo}.

The saddle-point equation for the action \eqref{action} at the critical point reads
\begin{equation}
\label{saddle_point}
\bigtriangleup m(\mathbf{x})=g \frac {d(d-2)}{4} m(\mathbf{x})^{\frac{d+2}{d-2} }
\end{equation}
with $m(\mathbf{x})=\braket{\phi(\mathbf{x})}$.

We will now specialize the 
study of the saddle-point 
approximation for the action~\eqref{action}  
for the physical dimensions
taking the %special 
values $d=3$, $4$, and $6$.
This is firstly due 
to their physical relevance, but 
also for the striking 
mathematical properties they display
only for these values of $d$. 
The peculiarity
amounts to the fact that they are the only
dimensions where the magnetization profiles
can be expressed as suitable powers of the 
Weierstrass elliptic function $\wp$
(see Appendix \ref{app:Wei} and \cite{NIST:DLMF} for reference). 
{Before moving to the different cases $d=3$, $4$, and $6$ 
in Secs. \ref{sub:Is}--\ref{sub:d6}, we discuss the different boundary conditions we are going to treat.} % in the next subsection \ref{sub:BC}.}

%{For $d=4$ the action \eqref{action} is the well-known $\phi^4$ theory. Such action can be generalized to a vector field $\vec{\phi}$ with $N$ components: the potential term becomes $||\vec{\phi}||^4$. In this case, there is}

%For the cases $d=3,4$, the action \ref{action} can be generalized to describe a vector field with $N$ components: In this case, the potential term becomes $||\phi(x)||^{2d/(d-2)}$, where $||\phi(x)||$

%{The action \eqref{action4} can be easily generalized to a vector field $\vec{\phi}$, with $N$ components: the potential becomes $||\vec{\phi}||^4$. In this case, there is}

\subsection{Boundary conditions}
\label{sub:BC}

{Let us start from the Ising model. We have different choices of boundary conditions on a slab; forcing the spins on the two boundaries to be aligned (fixed boundary conditions, FBCs) corresponds to diverging order parameter at the boundaries. One could have $++$ or $--$ FBCs (due to symmetry, we consider only the $++$ case). If the spins  on the two boundaries are antiparallel, then one has  $+-$ FBCs. }

{The corresponding, conformally invariant,  boundary condition in the field theory is $\braket{\phi}=\pm \infty$.
Of course, on a lattice the order parameter cannot 
diverge: the magnetization at the boundaries will be $\pm 1$. 
For the $++$ FBCs, the value at the center of the slab, however, will scale with the system size as $L^{-\Delta_{\phi}}$, so one can {\it rescale} the magnetization
in a way that it is constant at the center. In the thermodynamic limit, it then diverges at the slab edges. It is this rescaled form which is accessible via field theory. }

{Another way to understand it is the following: if the lattice magnetization profile were prolonged a few sites beyond the boundary, these singularities would appear. This concept is quantified by the extrapolation length $a_L$ 
\cite{Diehl1997}: if the boundaries where we fixed the spins are at $x=0$ and $x=L$, after we fit the resulting profile $m(x)$, we would see divergences at $m(-a_L)$ and at $m(L+a_L)$.} %Another way to understand how the divergences at the boundaries appear is to notice that the magnetization at the center of the slab decays as $L^{-\Delta_{\phi}}$: in order to compare profiles with different $L$, they need to be rescaled by a factor $L^{\Delta_{\phi}}$. 
{In the $L\rightarrow\infty$ limit, $a_L\rightarrow 0$ and the magnetization diverges at the boundaries.}

{One can also fix the spins on one boundary to the value $+1$ and the spins on the other boundary to the value $-1$ ($+-$ FBCs). The corresponding magnetization profile vanishes in the center, so either half of this profile could be obtained by fixing the spins on one boundary, and leaving the other boundary free (mixed boundary conditions, $+0$.)

{We clarify that in the following, in agreement with action (\ref{action}), we do not consider an external magnetic field, except for the one needed to fix the values of the spins at the boundaries. Therefore open boundary conditions (OBCs) are trivial, in the sense that the magnetization at the critical point has to vanish. In Appendix \ref{app00} we discuss the case of OBCs with a suitably chosen scaling magnetic field, providing an illustration of the usefulness of the calculation developed in this section and a further interesting perspective on the mathematical structure of the solutions.}
 %Possibly, one could fix a magnetic field and take the thermodynamic limit, or vice versa, and the order of limits is in general important. We will come back on the choice of the magnetic field $h(x)$ later in the text.}

{From the scaling dimension $\Delta_{\phi}$ of the order parameter, we can also obtain the behavior near the boundary (see Table \ref{bctab})}

\begin{table}[ht]
\centering
  \begin{tabular}{ c | c | c | c}
  \hline 
     BC & Lattice & $m(x\rightarrow 0)$ & $m(x\rightarrow L)$ \\
    \hline
    $++$ & $\uparrow \dots \uparrow$ & $x^{-\Delta_{\phi}}$   & $(L-x)^{-\Delta_{\phi}}$\\
    $+-$  & $\uparrow \dots \downarrow$ & $x^{-\Delta_{\phi}}$ & $-(L-x)^{-\Delta_{\phi}}$ \\
    $+0$  & $\uparrow \dots 0$ & $x^{-\Delta_{\phi}}$ & $L-x$ \\
    \hline
    \end{tabular}
  
\caption{Labels for the possible boundary conditions, with the corresponding spin configuration and behavior of the continuous profile near the boundaries.} \label{bctab}
\end{table}

{If we were to consider $O(N)$ models with $N>1$, there would be an additional degree of freedom in the choice of boundary conditions: the angle between the spins on the two boundaries. The magnetization becomes then a vector in the plane spanned by the boundary spins, and the saddle-point becomes a system of two equations. So, while the $+-$ solution is specific to the Ising model, the $++$ one is valid for any $O(N)$ model with parallel boundary spins.} {Here we focus on conformally invariant boundary conditions, which are homogeneous on the slab plates. Other boundary conditions are possible, such as the one considered in \cite{Panero2021} to enforce a topological excitation.}

\subsection{$\phi^4$ theory in four dimensions}
\label{sub:Is}

%The Landau-Ginzburg description 
%for systems with $O(N)$ symmetry 
{In four dimensions the action \eqref{action} %is the well-known $\phi^4$ theory:
reads}
\begin{equation}
 S=\int d^4x\left[ \frac 1 2\phi(-\bigtriangleup)\phi+\frac 1 2 \mu^2\phi^2+\frac{1}{2} g\, \phi^4 \right].
\label{action4}
\end{equation}
{A corresponding model in the lattice is of course the Ising model.} Since the theory is at the upper critical dimension, we proceed by writing the saddle-point equation. The saddle-point equation for the action at the critical point $\mu=0$ is
\begin{equation}\label{y4d}
\bigtriangleup m(\mathbf{x})=2\,g\,m(\mathbf{x})^3.
\end{equation}

We are interested in solving Eq. \eqref{y4d} in the case of a slab domain $[0,L]\times \mathbb{R}^3$. The magnetization $m(\mathbf{x})$ depends only on the transverse direction $x$, so the Laplacian $\bigtriangleup$ reduces to $\partial_x^2$. The different solutions for various boundary conditions will be analyzed and compared in the rest of the this section. In Sec.~\ref{MC} we will compare the analytical results so obtained with Monte Carlo simulations for the Ising model on a four-dimensional slab geometry with fixed boundary conditions. 

Notice that given a solution $m(x)$ of \eqref{y4d}
we can generate other solutions by translation
and rescaling ruled by the parameters $x_0$, $\beta$, and $\lambda$, all of them being possibly complex:
\begin{equation}{
    m(x)\rightarrow \beta\, m(\lambda\, (x-x_0)),\qquad g\rightarrow g\, \beta^{-\frac{4}{d-2}}\, \lambda^2.\label{y4dscal}}
\end{equation}
This allows us to change the domain or coupling constant
as desired; for convenience we set $g=1$.
In general such scaling will
alter the boundary conditions, however
scaling invariant conditions, i.e., $m=0$, $\infty$, are left unchanged. We remark that the scaling \eqref{y4dscal} 
also applies to other values of $d$ and it will thus be useful in the {next subsections}.

The general solution to Eq.~\eqref{y4d} {(with $g=1$)} is
\begin{equation}
{ m(x)= \lambda\, \wp_l(\lambda  (x-x_0))^{1/2},}
\end{equation}
where $\wp_l$ is the Weierstrass $\wp$ function with invariants $(g_2,g_3)=(1,0)$. This is the so-called lemniscatic case. When it is written in terms of the complex variable $z$ as $\wp_l(z)$, it corresponds in the complex plane to a square lattice, with a real half-period $\omega_l=\Gamma(\frac 1 4)^2/(4\sqrt{\pi})$. 

The Weierstrass elliptic $\wp$, whose basic properties are refreshed in Appendix \ref{app:Wei}, is a doubly periodic  function. Within one of its domains in the complex plane, it has a double pole and two zeros possibly coinciding (as here): taking segments with poles and/or zeros as their endpoints gives solutions with various boundary conditions, {as can be seen in Fig. \ref{fig1}}.
Along the segments the function has to be real or have a constant phase (which can removed)
in order to be interpreted as magnetization profiles.
%To see the corresponding boundary conditions in the lattice model, let us focus on the $N=1$ case, the Ising model. 

The corresponding (unscaled) solutions are
\begin{align}\label{bc}
& m_{++}(x)= \wp_l(x)^{1/2}, & x\in(0,2\,\omega_l), \nonumber& \\
& m_{+0}(x)= \wp_l(x\,e^{i \pi/4})^{1/2}
e^{i \pi/4}, &x\in(0,\sqrt{2}\,\omega_l]. &
\end{align}

\begin{figure}
\centering
\includegraphics[width=.38\textwidth]{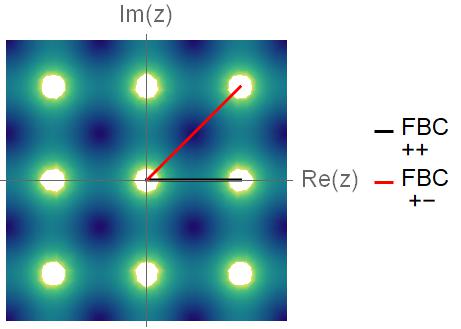}
\includegraphics[width=.325\textwidth]{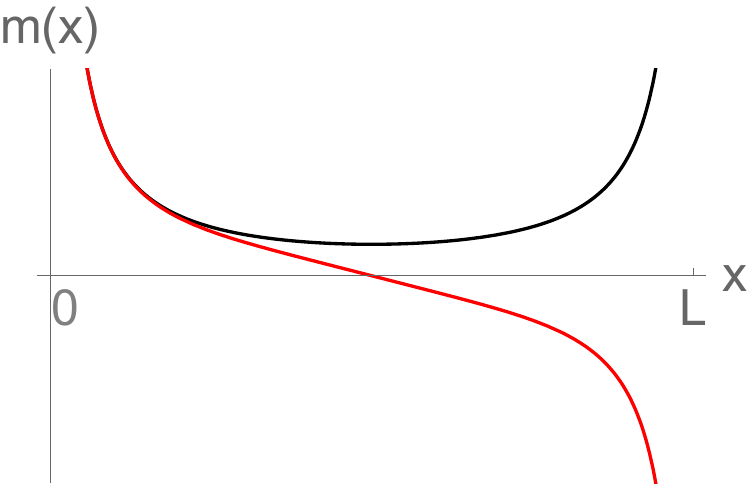}
\caption{$d=4$ case. Above: modulus of the square root of the lemniscatic elliptic function, poles are white and zeros blue. Below: magnetization profiles for the different boundary conditions discussed in the text.}
\label{fig1}
\end{figure}
%{The OBC solution warrants a few comments. The solution of the saddle-point equation \ref{y4d} with vanishing boundary condition is simply $m(x)=0$ throughout the slab. The solution we have provided is actually the solution to a saddle-point equation with the opposite sign:}
%When considering OBC, we have ignored the trivial solution $m(x)=0$ throughout the slab. Practically, $m_{00}(x)$ could be obtained by adding a small magnetic field in the center of the slab, to break the symmetry. 
By extending the $m_{+0}(x)$ solution
to the interval $x\in(0,2\sqrt{2}\omega_L)$
we indeed generate a solution $m_{+-}$ connecting 
the + and $-$ boundary states. This reflects the
$\mathbb{Z}_2$ symmetry of the model as it is apparent in the saddle-point equation \eqref{y4d} with $m(x)\rightarrow-m(x)$.
%{The $00$ solution warrants some comments. Joining two zeros of the $\wp_l(z)$ we obtain a profile which satisfies the saddle-point equation with opposite sign:}

%To get these profiles, we have not needed to specify the model, so they are valid for any system with $O(N)$ symmetry, in particular the Ising and XY models. Solutions with opposite boundary conditions (spins fixed up on one side and down on the other), however, are model-dependent.

\subsection{$\phi^6$ theory in three dimensions}
\label{sub:d3}
A $\phi^6$ theory has upper critical dimension $d_c=3$. The steps of the previous section can be repeated in this case. We start from the Landau-Ginzburg action at the critical point, where the couplings of the $\phi^2$ and $\phi^4$ terms vanish:
\begin{equation} S=\int d^3x\left[ -\frac 1 2\phi(-\bigtriangleup)\phi +\frac{g}{8}\,\phi^{6} \right].
\end{equation}
On the slab geometry $[0,L]\times \mathbb{R}^2$, we get the saddle-point equation
\begin{equation}
m''(x)=\frac 3 4\,g\, m^{5}, 
\end{equation}
where again due to the scaling property~\eqref{y4dscal} we set $g=1$. The solution is
\begin{equation}\label{pe3}
{m(x)=\lambda^{1/2}\wp_e (i\, \lambda \,(x-x_0) )^{-1/2}. }
\end{equation}
In the present case, the invariants of the Weierstrass function $\wp_e$ are $(g_2,g_3)=(0,1)$, from which its real half-period is  $\omega_e=\Gamma(\frac 1 3)^3 / (4\pi)$. 

The above case $\wp_e$ of the Weierstrass function is the so-called equiharmonic case, where the lattice
used to define the elliptic function is the regular triangular tiling of the plane; see 
Fig. \ref{fig2}.

Remarkably, the solution for $d=6$, presented in the next subsection, will turn out to be dual to this one.
Again, looking at the poles and zeros of $m(z)$, we 
obtain solutions for the possible boundary conditions:
\begin{align}
& m_{++}(x)= \wp_e(\omega_e+i\,x)^{-1/2}, & x\in(-\frac{\omega_e}{\sqrt{3}},\frac{\omega_e}{\sqrt{3}}),
&\nonumber\\
&m_{+0}(x)= \wp_e(x\,e^{i \pi/6})^{-1/2}
e^{-i \pi/6}, & x\in(-\frac{2\omega_e}{\sqrt{3}},0].&
\end{align}

These solutions could also be expressed through Jacobi elliptic functions, as was done for the $++$ solution in \citep{Borjan1998}, through Fisher-De Gennes theory \citep{Fisher1990}. Again due to $\mathbb{Z}_2$ 
invariance, reflected in the $m(x)\rightarrow -m(x)$
symmetry in the saddle-point equation, the 
$m_{+0}(x)$ can be extended further in the interval
$x\in(-\frac{2}{\sqrt{3}}\omega_e,
\frac{2}{\sqrt{3}}\omega_e)$ yielding a $m_{+-}(x)$ profile. 

\begin{figure}
\centering
\includegraphics[width=.38\textwidth]{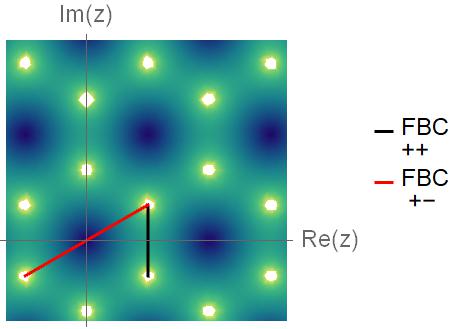}
\includegraphics[width=.325\textwidth]{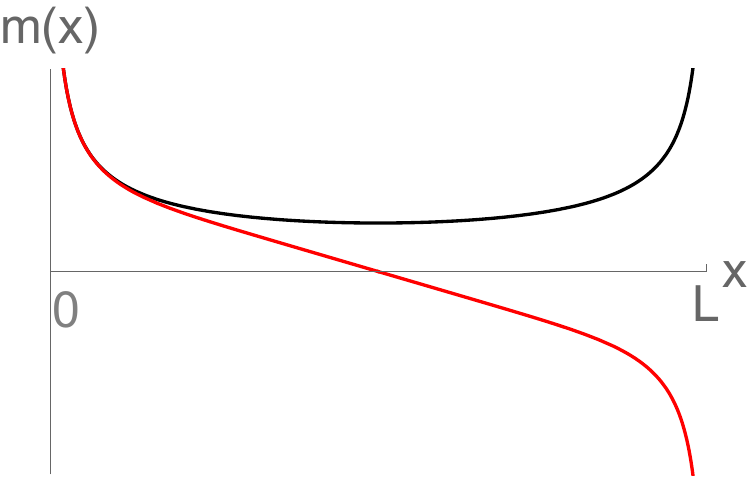}
\caption{As in Fig. \ref{fig1}, but here for $d=6$, the modulus of $m(z)$ in the complex plane, with the solutions for the different boundary conditions.}\label{fig2}
\end{figure}

\subsection{$\phi^3$ theory in six dimensions}
\label{sub:d6}

%Finally we turn to the $\phi^3$ theory in $d=6$, whose action,
{In $d=6$, we  consider an action with cubic potential:}

\begin{equation} S=\int d^6x\left[ \frac 1 2\phi(-\bigtriangleup)\phi +2\,g\,\phi^{3} \right],
\end{equation}
which is used to describe percolation at the upper critical dimension \cite{amit1977,Bonfirm1981}.
This {action is not obtained directly by plugging $d=6$ into \eqref{action}, so it} differs from the previously considered $d=3$, $4$ cases;  {since the potential is odd,} it lacks $\mathbb{Z}_2$ symmetry. This implies the absence of opposing ($+-$) boundary conditions, which becomes clear when one thinks of the possible boundary conditions for percolation: one can either force a boundary to belong to the percolating cluster ($+$), or leave it free, but opposing boundaries no longer make sense. However, we can still find a solution  connecting the $+$ and $0$
boundary states.
The saddle-point equation now is
\begin{equation}
m''(x)=6\,g\,m(x)^{2}.
\end{equation}
The general solution is (taking $g=1$ as in the previous cases) the function $\wp_e$: 
\begin{equation}\label{pe6}
{m(x)=\lambda^2\wp_e (\lambda \,(x-x_0)),}
\end{equation}
meaning that the profiles are just the square of the reciprocal of the $\phi^6$ results, as evidenced in Fig. \ref{fig3}. 
Now the double pole in the origin directly 
yields the expected divergence for
the $m_{++}$ order parameter profile,
since the dimension of the field 
is $\frac{d-2}{2}=2$. The profiles are:
\begin{align}
&m_{++}(x)= \wp_e(x),& x\in(0,2\omega_e), &&\nonumber\\
&m_{+0}(x)= \wp_e(x\,e^{i \pi/6}) e^{i \pi/3},
&x\in(0,\frac{2\omega_e}{\sqrt{3}}]. &&
\end{align}

\begin{figure}
\centering
\includegraphics[width=.38\textwidth]{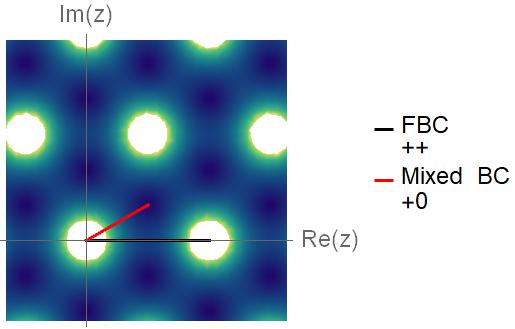}
\includegraphics[width=.325\textwidth]{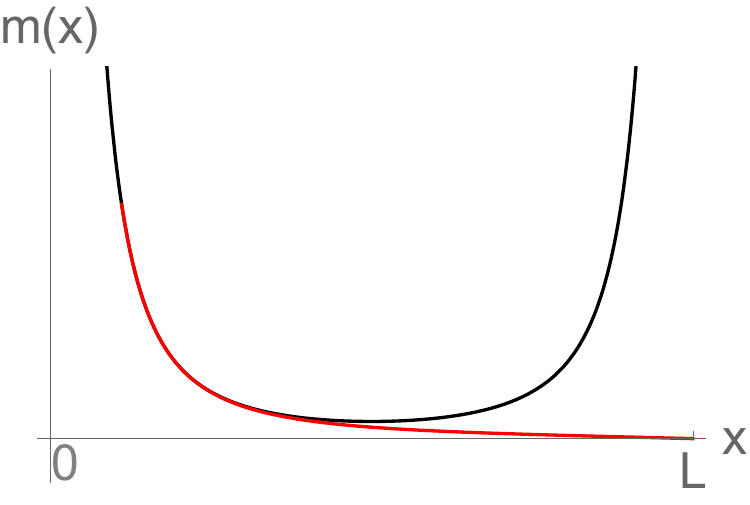}
\caption{The zero and pole structure for $d=6$ is inverted with respected to Fig. \ref{fig2}. What would be the $+-$ profile now ends at $0$, giving the mixed boundary condition $+0$.}\label{fig3}
\end{figure}

\section{%Reshaping the saddle-point equation
The Yamabe equation}
\label{Yamabe}

In the following we provide the solution to the saddle-point
equation for $\phi^{2m}$ theories, valid for %other 
critical dimensions 
$d_c=\frac{2m}{m-1}$ 
(in this general case, they will not
possess the analytic properties displayed by
the $d=3$, $4$, $6$ cases discussed above). 

%In order to do straightforwardly make this computation, and to 
%However, before doing that we will express the
%saddle-point equations in a slightly more general %way, that will help us to 
In order to introduce a geometric interpretation of the critical magnetization profile, 
%main idea of the paper. 
we notice that it is possible to re-write the saddle-point equation in a convenient form. Since the scaling dimension of the order parameter field is $\Delta_{\phi}=\frac{d-2}2$, 
%it seems reasonable to 
we introduce a function $\gamma(\mathbf{x})$ which acts as a point-dependent scale. % with the power exponent given by (minus) the scaling dimension of the field operator $\phi$. 
We then write
\begin{equation}\label{reshape}
m(\mathbf{x})=\alpha\gamma(\mathbf{x})^{-\Delta_\phi}=
\alpha\gamma(\mathbf{x})^{-\frac{d-2}2},
\end{equation}
where $\alpha$ is a constant, to be fixed.

The main point of this section, as well as the main result of the paper, is that the solution 
$m(\mathbf{x})$ of the saddle-point equation \eqref{saddle_point}, once rescaled as in \eqref{reshape}, can be written as 
\begin{equation}
(-\bigtriangleup)\gamma(\mathbf{x})^{-\frac{d-2}2}=-\frac{d(d-2)} 4 \gamma (\mathbf{x})^{-\frac{d+2}{2}},
\label{YAM}
\end{equation}
where $\alpha$ in \eqref{reshape} is chosen to be $\alpha=g^{(2-d)/4}$ so that the right-hand side has the correct coefficient to obtain what is known as the Yamabe equation. Equation \eqref{YAM} can be also recast in the form
\begin{equation}\label{YAM_L}
    1-|\vv{\nabla} \gamma(\mathbf{x})|^2+\frac{2}{d} 
    \gamma(\mathbf{x})\bigtriangleup\!\gamma(\mathbf{x})=0.
\end{equation}

The Yamabe equation has the geometric interpretation of a constraint on the curvature of a manifold.
This equation can be viewed as a special case of the more general  {\it fractional} Yamabe equation, which contains a fractional Laplacian, allowing it to describe theories with nonvanishing anomalous dimension as done in \cite{gori}. In the following, whenever we mention the Yamabe equation, we are referring to the integer one, with standard Laplacian, as first introduced in \cite{Yamabe1960}
.%Later ???, we will introduce a variant of the Yamabe equation that contains a fractional Laplacian, which is therefore called When ambiguity may arise, we will then refer to \eqref{YAM} as the {\it integer} Yamabe equation, to emphasize that it contains the standard Laplacian.

%In the following, we wiil refer to the yamabe equation (\ref{YAM}) also to the {\it integer} Yamabe equation, to emphasize that the Laplacian in it is not fractional, as opposed to the {\it fractional} Yamabe equation, where the Laplacian is fractional.
To understand how a geometric problem emerges from one-point functions of a bounded system at criticality, we 
discuss the connection of the obtained results with the ``uniformization'' hypothesis put forward in \cite{gori}. %This apparently unrelated meaning can be linked back to bounded critical systems through the hypothesis of uniformization.

\subsection{Uniformization hypothesis and the Yamabe equation}

The main property a system typically gains at its critical points is conformal invariance \cite{Polyakov1970,Polyakov1974}. Heuristically, this means that every point and every region of the system look the same. Introducing a boundary clearly breaks this property. The question addressed in \cite{gori} is then: is there a way to recover it? If a metric is introduced which sets the boundary at an infinite distance, then we would no longer have a distinction between points close to the boundary and points deep in the bulk. The only changes of the euclidean metric that we can allow are conformal transformations, since the system must still be locally euclidean. This means that the choice of metric reduces to the choice of a function $\gamma(\mathbf{x})$ which sets a local scale:
\begin{equation}
\delta_{ij}\rightarrow g_{ij}=\frac{\delta_{ij}}{\gamma(\mathbf{x})^2},
\end{equation}
$\delta_{ij}$ being the flat metric, i.e.,  the identity matrix, and { $i,j=1,\ldots,d$.}

Constraints on the function $\gamma$ have to be imposed. Since we have a curved space, we should look into the various quantities that describe its curvature, the most obvious one being the Ricci scalar curvature. The main hypothesis in \cite{gori} is that the metric must make a bounded critical system as uniform as possible: this means making the scalar curvature constant. This curvature would have to be negative, since spaces with positive curvature, like spheres, lack boundaries. The simplest examples of space with constant negative curvature are the Poincaré half plane and disk models. 

%\subsection{Yamabe equation}

Starting from the metric with an unknown $\gamma(x)$, one can compute the Christoffel symbols, from which one gets the Ricci tensor, and finally the Ricci scalar:

\begin{equation}
R=\sum_{i,j=1}^d R_{ij}g^{ij}=\kappa, \qquad \kappa<0.
\end{equation}
Without losing generality, we can set $\kappa=-1$. We can now write this condition as an equation for the factor $\gamma(\mathbf{x})$, getting \eqref{YAM}, 
%
%\begin{equation}\label{yamabe}
%(-\bigtriangleup) \gamma %(\mathbf{x})^{-\frac{d-2}{2} %}=-\frac{d(d-2)}{4}\gamma %(\mathbf{x})^{-\frac{d+2}{2}},
%\end{equation}
supplemented by the condition $\gamma(\mathbf{x})=0$ at the boundaries
of the domain $\Omega$. This is enough to get solutions which, close to the boundary, are proportional to the euclidean distance from it. The distance from any point to the boundary, computed with this metric, is therefore infinite, as desired. 

To frame it more generally, the Yamabe problem consists in finding a metric in the conformal class of another given metric for a smooth Riemann space, that makes the scalar curvature constant \cite{Lee1987}. A solution has been proven to exist for any such space, provided it is compact. 

The two-dimensional limit of \eqref{YAM} gives the Liouville equation
\begin{equation}
(-\bigtriangleup) \log \gamma (\mathbf{x})=-\gamma(\mathbf{x})^{-2}.
\label{Liouville}
\end{equation}

The Yamabe equation is a nonlinear differential equation.%, analytic solutions are not trivial to find. 
 There are a few cases where the Yamabe equation has simple solutions: e.g., for a ball of radius $R$ in any dimension, one finds $\gamma=\frac{R^2-|\mathbf{x}|^2}{2R}$ \cite{MarGonzalez2010,gori}; the case of the slab geometry will be treated in the next section. 
%By sending
%the radius $R$ to infinty the solution
%for the half hyperspace ($x_d>0$) is also obtained 
%$\gamma=x_d$. 

One-point functions and the scale factor transform similarly under a dilation of the system $\Omega \rightarrow \lambda \Omega$:
\begin{equation}
\braket{\phi_{\lambda\Omega}(\lambda \mathbf{x})}=\lambda^{-\Delta_{\phi}}\braket{\phi_{\Omega}(\mathbf{x})}, \qquad \gamma_{\lambda\Omega}(\lambda \mathbf{x})=\lambda \gamma_{\Omega}(\mathbf{x}),
\end{equation}
where $\Delta_{\phi}$ is the scaling dimension of the field. %This suggests that, 
Therefore, once $\gamma(\mathbf{x})$ is known, all one-point functions are determined up to a constant ${\cal C}$:
\begin{equation}\label{1p}
\braket{\phi(\mathbf{x})}=
\frac{{\cal C}}{\gamma(\mathbf{x})^{\Delta_{\phi}}}.
\end{equation}
As an example, for a half space in any dimension, {with $x_1>0$ and $\{x_2, \ldots, x_d\} \in \mathbb{R}^{d-1}$ }, we have $\gamma(\mathbf{x})=x_1$, so
\begin{equation}
\braket{\phi(\mathbf{x})}=\frac{{\cal C}}{x_1^{\Delta_{\phi}}},
\end{equation}
reproducing a standard result of boundary conformal field theory \cite{CardyBCFT}.

The introduction of $\gamma(\mathbf{x})$ in \eqref{reshape} should now be clearer: the mean-field equation for a (multi-)critical of the $O(N)$ model is equivalent to the Yamabe equation at the corresponding upper critical dimension. Once the scale factor $\gamma(\mathbf{x})$ is obtained by solving the Yamabe equation, the magnetization is recovered through \eqref{1p}.

\section{Analytical results for the slab geometry}
\label{analytics}

For the slab geometry we can actually obtain
the solution for general dimension $d$ in implicit form. 
We denote the solution of the Yamabe equation \eqref{YAM} with $\gamma_d(x)$ to emphasize the dependence on the dimension $d$. For convenience we assume $x\in[0,2]$ (i.e., $L=2$), so that the center of the slab is in $x=1$. 

For $x\in [0,1]$ and $++$ FBC, the relation between $x$ and $\gamma_d$ is given in terms of the ${}_2F_1$ hypergeometric function. From the Yamabe equation one straightforwardly finds 
\begin{equation}
    x(\gamma_d) =  {}_2F_1\left( \frac 1d,\frac 12; 1+\frac 1d; \left(\frac{\gamma_d}{\gamma_m}\right)^d\right)\,\gamma_d\label{gammad}.
\end{equation}
By reflecting $x$ around $1$: $x\rightarrow 2-x$ the other branch is obtained. The constant
\begin{equation}
    \gamma_m = \frac{\Gamma\left(\frac 12 + \frac 1d\right)}{\sqrt{\pi}\,
    \Gamma\left(1 + \frac 1d\right)} 
\end{equation}
is the (maximum) value acquired by 
the conformal factor at the center of the slab. Equation \eqref{gammad} is valid for {\it any} $d \ge 2$, including the cases $d=4,6,3,2$ and also real values of $d$. Notice that a slab configuration cannot be defined for $d<2$. The formulas for other boundary conditions are written below.

We pause here to comment that one sees a (minor) advantage of using the Yamabe equation in the form 
\eqref{YAM_L} instead of using directly the saddle-point equation \eqref{saddle_point}. The latter, 
when written in the slab geometry, gives rise to the so called Emden-Fowler equation \cite{Polyanin}. The latter, for a function $m(x)$, reads in its canonical form $m''(x)=A\,x^{\cal N} m(x)^{\cal M}$ 
\cite{Polyanin}, with our case corresponding evidently 
to 
${\cal N}=0$ and ${\cal M}=(d+2)/(d-2)$. The case ${\cal N}=0$ can be solved by quadrature \cite{Polyanin}, 
writing the solution in term of an integral and for certain values of ${\cal M}$ the corresponding analytical expressions are tabulated \cite{Polyanin}. However, 
solving the Yamabe equation \eqref{YAM_L} and using the procedure prescribed in ordinary differential equations textbooks (or directly \verb|Mathematica|) one finds that the solution is given by the inverse of the hypergeometric function ${}_2F_1$ for any $d$.
%Of course, we do not say that one cannot solve for the magnetization using the saddle-point Ginzburg-Landau form 
%\eqref{saddle_point}, since the latter and the Yamabe are just mapped one in the other as shown in the previous Section. We just say that 
Therefore, rewriting $m$ in terms of $\gamma$ via the equation \eqref{reshape} may 
also help to find easier analytical solutions, as the case of the slab geometry shows. In Appendix \ref{app00}, we discuss how to relate the result \eqref{gammad} to different boundary conditions, including open boundary conditions with an external magnetic field.

We can see what the general result \eqref{gammad} simplifies to when we substitute  $d=4,3,6,2$:
\begin{itemize}
    \item[$d=4$:] the inverse of \eqref{gammad} is $\wp_l(x)^{-1/2}$, the lemniscatic Weierstrass function used in \eqref{y4d}, see Appendix \ref{appB}.
    \item[$d=6$:] the inverse is the equiharmonic elliptic function $\wp_e(x)$ as seen in \eqref{pe6}.
    \item[$d=3$:] this case is dual to the $d=6$ case, since the solution here is simply the square root of the reciprocal of the previous solution, {after an appropriate translation; this is clarified in Appendix \ref{app00}}.
    \item[$d=2$:] the value $d=2$ cannot be directly chosen in the Yamabe equation \eqref{YAM}: a limit has to be performed, yielding the Liouville equation 
    \eqref{Liouville}. One may then assume that plugging $d=2$ into \eqref{gammad} will give an incorrect result. Surprisingly, that is not the case: for $d=2$, the function $_2F_1(\frac 1 2,\frac 1 2; \frac 3 2; \gamma^2\frac{\pi^2}{4})\gamma$ reduces to the inverse sine: indeed, $\gamma_2(x)=\frac 2{\pi} \sin(\pi x/2)$. 
\end{itemize}
{This agrees with and  generalizes previous results:} profiles for $++$ and $+-$ boundary conditions were found for $d=4$ in \cite{Krech1997}, and for $d=2$ in \cite{Carlon1998}; the $++$ profiles for $d=2,3,4$ are also found in \cite{Borjan1998}. In both \cite{Krech1997} and \cite{Borjan1998} Jacobi functions were used. Switching to Weierstrass functions allowed us in the present paper to write profiles for various dimensions in a %consistent and 
compact way, retrieving the previously listed known  results, finding new results for all the conformally invariant boundary conditions. We also obtain new results for all the critical magnetization profiles in $d=6$. 

\section{Solution of lattice mean-field equations
%of the Ising model
}
\label{MF}

\begin{figure}
\centering
\includegraphics[scale=.5]{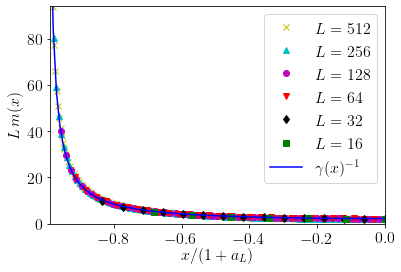}
\caption{Collapse of the mean-field magnetization profiles. Since they are symmetric around the center of the slab, we plot only the left half. Each set of magnetization values has been rescaled by a factor $L$, while the $x$ coordinate has been rescaled through its extrapolation length.}\label{iterativo}
\end{figure}
%To see more straightforwardly how lattice profiles can be accurately described by a continuous function from field theory, 
In this section we discuss the solution of the mean-field equations for the Ising model in a slab geometry directly on the lattice. The goal is to see how lattice profiles can be accurately described by a continuous function from field theory in the thermodynamic limit. For this reason in this section we take the slab coordinate $x$ to be an integer 
variable, going from $1$ to $L$, and then take $L \to \infty$. In this calculation it is not necessary to specify the number of sites in the directions perpendicular to $x$, since we can assume that the probability distribution function $P$ of any spin depends only on its transverse coordinate $x$. For this reason we 
put
\begin{equation}
P(s_i)=\frac{1+m_x}2 \delta_{s_i,1}+\frac{1-m_x}2 \delta_{s_i,-1},
\end{equation}
where $s_i=\pm 1$ is the Ising variable in the site $i$ having $x$ as coordinate along the direction of the slab.

We can find the discrete magnetization profile 
$m_x=\langle s_i \rangle$ minimizing the free energy. The mean-field equation \cite{LeBellac} is 
\begin{equation}
m_x=\tanh \left(\beta \sum_j \braket{s_j}\right).
\end{equation}
This is the same equation used to find the inverse critical temperature $\beta_c$ in the bulk, i.e. in the case of no  boundaries. Having set the  coupling $J$ to $1$ for convenience, it is 
$\beta_c=1/2d$. Of course, it differs from the critical temperature of the actual Ising lattice, see the next Section. 

In a four dimensional slab with $d=4$, this becomes
\begin{equation}
m_x=\tanh \left[\beta(6 m_x+m_{x+1}+m_{x-1}) \right].
\end{equation}
For the case of FBC $m_1=m_L=1$, this equation can be solved iteratively until the difference between the two sides is smaller than a fixed threshold.

The value of the magnetization at the center of the slab decays, as the system size increases, as $1/L$. Since profiles for different sizes must have the same functional shape at the critical point, we rescale them by multiplying each profile by its corresponding size $L$. This also clarifies the correspondence between fixing the boundary spins to $+1$ in the lattice model and diverging boundary conditions in the field theory. As the size increases, the rescaled boundary magnetization grows proportional to $L$. 

At the same time, as $L$ increases, the extrapolation length $a_L$ \cite{Diehl1997} decreases, meaning that the point where the profile diverges gets closer to the lattice boundaries. In the limit $L\rightarrow \infty$, $a_L\rightarrow 0$ and {the rescaled}  $m(0)=m(L)\rightarrow \infty$.

Once the rescaling is done, we clearly see a collapse of the various profiles, showing that we are at the critical point. The corresponding results are in Fig. \ref{iterativo}. For large $L$, the extrapolation length vanishes and the mean-field profile coincides with the saddle-point solution. {We also studied the solutions of the lattice mean-field equations for $+0$ boundary conditions, finding a similarly good agreement between them and the solution of the corresponding Yamabe equation.}

%A test of \eqref{2p}, on the other hand, cannot be performed at the mean field level, since correlations factorize: we trivially obtain $\mathcal{F}\left(\mathcal{D}_g(\mathbf{x},\mathbf{y}) \right)=1$. This means that a study of two-point correlation must be performed below the upper critical dimension, which requires tweaking the Yamabe equation.
%We see a clear collapse away from the boundary, while for larger sizes the points near the center move away from the center and from a field theoretical solution.

\section{Monte Carlo simulation of the $4d$ Ising model on a slab}
\label{MC}

The previous predictions have been obtained by performing the saddle-point approximation on the action \eqref{action}. 
It is very well known that in four dimensions for the $\phi^4$ theory, and in general at the upper critical dimension, the critical exponents are the mean-field ones \cite{Cardy1996,Mussardo,Amit,LeBellac}. This is routinely exploited in conformal bootstrap calculations for the bulk geometry, where in $d=4$ the critical theory is Gaussian \cite{RychkovEPFL}. However, in a bounded domain---to the best of our knowledge---there is no proof that the critical magnetization profile in the thermodynamic limit given by the saddle-point 
approximation is exact in $d=4$ for the $\phi^4$ theory, although it is expected. For this reason we decided to numerically test the saddle-point findings by Monte Carlo simulations and validate our predictions 
via numerical experiments. In order to obtain a numerical check of the magnetization profiles, we will concentrate 
on the Ising model on the slab geometry with FBC. 
Notice that explicit numerical investigations for
 high dimensions of phenomena arising from
inhomogeneities are rather sparse especially if compared to two-dimensional systems. Three dimensional models
did receive of course attention, and basic predictions from scale and conformal invariance 
have recently been tested~\cite{Gori2015,Cosme2015}.
%This will entail the simulation
%with Monte Carlo methods of a $\mathbf{Z}_2$ invariant
%model, that is the Ising model for $d=4$ in
%a slab geometry.

We performed Monte Carlo simulations of the Ising model at its upper critical dimension, $4$, in a slab of sizes $L$ in the transverse direction, and $4L$ in the other three directions; $L$ ranges from $16$ to $56$. The FBC are implemented by fixing the spins to the same ($+1$) or opposite ($\pm1$) on the two faces in the transverse direction, while the other directions have periodic boundary conditions. We can then expect the $++$ and $+-$ magnetization profiles from \eqref{bc}. The critical inverse temperature $\beta_c$ is taken to be $\beta_c=0.1496927$ from \cite{Luijten1997}, see also \cite{Bittner,Lundow2009}.

The simulation uses the standard Metropolis algorithm, whose moves are single spin flips. After a thermalization time, we measured the average magnetization and local energy along hyperplanes parallel to the boundaries. As explained in Secs. \ref{magprofiles} and \ref{MF}, since the spins are fixed at the edge faces 
and one is at the bulk critical temperature, the magnetization profile will start from the value $1$ at one boundary, will decrease as one approaches the center, should reach a value of order of $1/L$, and then rise again up to $1$ in the other slab boundary. One can rescale the numerical data in a way that the value of the magnetization at the center is constant, meaning that the boundary magnetization increases with $L$.

\begin{figure}
\centering
\includegraphics[scale=.5]{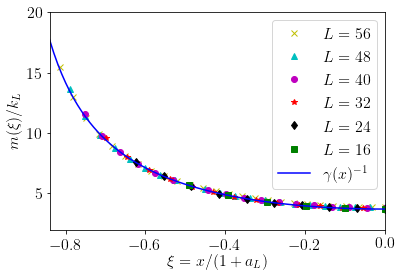}%{isingSENZAL56.png}

\caption{Collapse of the critical magnetization profiles in the four-dimensional Ising model for slabs of different sizes {with ++ boundary conditions}. Again, we plot only half the profile. Each set of points has been rescaled with a multiplicative constant and its extrapolation length, obtained from the fit \ref{fit_}.}
\label{Ising_fit}
\end{figure}

\begin{table}[h]\centering
\begin{tabular}{  c  c  }
 \hline
 $L$ \qquad & \qquad $a_L$  \\
 \hline 
 16 \qquad & \qquad 0.2465   \\
 24 \qquad & \qquad 0.1724\\
 32 \qquad & \qquad 0.1314\\
 40 \qquad & \qquad 0.1070\\
 48 \qquad & \qquad 0.0884\\
 56 \qquad & \qquad 0.0404 \\
  \hline
\end{tabular}

\caption{Decreasing extrapolation lengths as the size increases ($++$BC).}

\end{table}
\begin{figure}[h]
\centering 
\includegraphics[scale=.5]{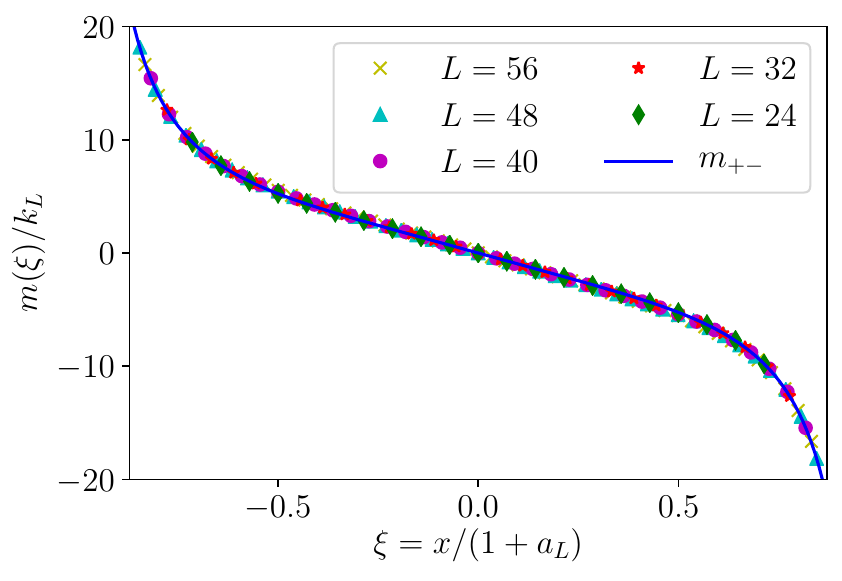}%{isingSENZAL56.png}

\caption{{Collapse of the critical magnetization profiles in the four-dimensional Ising model for $+-$ boundary conditions.}}
\label{fitpm}
\end{figure}

The magnetization data are then fitted with
\begin{equation}
m(x)= k_L \left[\gamma\left(\frac x {1+a_L}\right)
\label{fit_}\right]^{-1},
\end{equation}
% the $\xi^2$ per degree of freedom for different sizes: we see it approaching zero for larger systems, meaning that the magnetization in an $L=\infty$ slab would be described exactly by the saddle-point profile.
where {$k_L$} is a multiplicative constant roughly proportional to the system size, and $a_L$ is the extrapolation length, with the index $_L$ denoting the dependence on the size. The extrapolation length shrinks the lattice profile so that the divergence of the order parameter would appear a few sites beyond the edges \cite{Cardy1996,Diehl1997}. We see that it becomes smaller as the size grows, meaning that larger sizes describe a larger portion of the saddle-point profile. 
In Fig. \ref{Ising_fit} and \ref{fitpm}, we plot the magnetization profiles obtained for different sizes, compared with the prediction from \eqref{1p}. The magnetization $m(x)$ for different sizes are plotted as functions of the respective rescaled variable $\xi=x/(1+a_L)$, which highlights the collapse. {Simulations with $+0$ boundary conditions would look like half of the profile shown in Fig. \ref{fitpm}.}

Despite qualitative agreement between the prediction and the numerical data, the distance between raw data points and the theoretical curve is larger than the estimated numerical error (in Figs.  \ref{Ising_fit} and {\ref{fitpm}} smaller than the point sizes). This is to be expected: while \eqref{1p} gives the correct mean-field behavior, observables at the upper critical dimension also include some logarithmic corrections. %that cannot be captured by our theory. 
To account for them, we proceed as in \cite{Luijten1997}. At the critical temperature, the main finite size corrections to the bulk magnetization take the form
\begin{equation}
m%_{\mathrm bulk}%(L)
=c\frac{(\ln L)^{1/4}}L\sqrt{b_0+\frac {b_1}{\ln L}+\frac{b_2}{(\ln L)^2}},
\label{Luijten}
\end{equation}
where terms with higher powers of $1/\ln L$ have been neglected. In our case, the magnetization is a function of $x$ (or rather of the rescaled coordinate $\xi$), and therefore so are $b_0$, $b_1$ and $b_2$. %To show that this is the correct form of the logarithmic correction, 
For every $\xi$ we find the values of $b_0(\xi)$, $b_1(\xi)$ and $b_2(\xi)$ which best fit $m(\xi,L)$, seen as a function of $L=\{16,24,32,40,48,56\}$ \footnote{Notice that in \cite{Luijten1997} the (bulk) $4D$ Ising model at criticality is studied for linear sizes $L'$ from $2$ up to $48$ (to match with our notations, the hypercube in \cite{Luijten1997} is 
$L' \times L' \times L' \times L'$, while our slab is $4L \times 4L \times 4L \times L$).}. 

In order to compare with the analytical prediction {\eqref{bc} for $++$ BC}, the constant $c$ in (\ref{Luijten}) is chosen so that $\sqrt{b_0(0)}=m(0)$. We then plot, in Fig. \ref{a0}, the ratio $\sqrt{b_0(\xi)}/m(\xi)$, and we see that it remains close to 1. As we approach the boundaries, the value of the first logarithmic correction $b_1$ grows, explaining the deviation of the ratio from 1. 
%Nevertheless, for large enough system size, the term containing $b_1$ will vanish, recovering the saddle-point profile. 
Larger values of $b_1$ simply imply that it is numerically harder to measure the values predicted by (\ref{1p}) near the boundaries, as it requires simulating even larger systems.

\begin{figure}
\centering
\includegraphics[scale=.5]{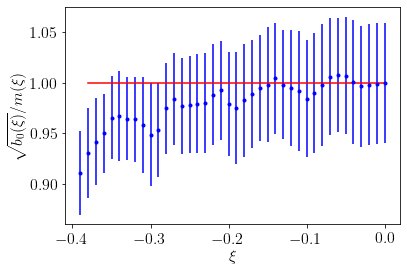}%{a0.png}
\caption{Ratio of the square root of the fit parameter $b_0$ and the expected magnetization ($++$BC). The boundary $x=-1$ correspond to $\xi=-1/(1+a_L)$, and the first few points close to the boundary must be discarded, so the plot starts from $\xi=-0.39$.}\label{a0}
\end{figure}

From $\xi=-1$ to $\xi=-0.4$ we decided not to plot the ratio $\sqrt{b_0(\xi)}/m(\xi)$ since, due to the values of the extrapolation length, for these $\xi$'s we do not have points for each of the considered values of $L$. We remind that we are in the left part of the slab geometry, $\xi=0$ corresponding to the center and $\xi=-1$ to the left edge (we discretize $\xi$ in steps of $0.01$). We notice that by fitting the points from the magnetization using (\ref{Luijten}) without the logarithmic corrections one obtains a clearly worse estimate of the $\chi^2$, confirming the validity of the fitting function (\ref{Luijten}). Moreover, for any $\xi$ we find $b_0>b_1/\ln L_{\mathrm max}>b_2/(\ln L_{\mathrm max})^2$, where $L_{\mathrm max}=56$ is the maximum value of $L$ we were able to simulate. This shows that each successive term is a smaller correction to the infinite-size term, proportional to $b_0$.
The standard deviations $\sigma$ in Fig. \ref{a0} are obtained by fitting the magnetization points without 
weighting them with their respective errors. The point for $\xi=0$ 
has ratio $\sqrt{b_0(\xi=0)}/m(0)=1$ by construction, and with standard deviation --- determined as explained above --- $\sigma=0.04$.
The final result, depicted in Fig. \ref{a0}, 
shows that $\sqrt{b_0(\xi)}$ is compatible with 
$m(\xi)$ within $\sigma$ for $-0.32 \le \xi \le 0 $ and within $2\sigma$ for $-0.39 \le \xi \le -0.33$. To obtain more data in the range $-1 < \xi \le -0.4$ one should have larger sizes. {For the $+-$ data reported in Fig. \ref{fitpm} similar results have been obtained.}

The conclusion is that for the data we have, the analytical predictions is in agreement with the Monte Carlo numerical results within $2\sigma$.

%
%
%
%To see more directly how the saddle-point solution %coincides with the mean-field lattice profile, we can %solve the Ising model in a slab in the mean-field %approximation, where logarithmic correction do not appear.

\section{Conclusions}
\label{concl}

In this paper we studied the magnetization profiles %of $O(N)$ models at their upper critical dimensions 
{of models with marginal interaction} for different (scale invariant) boundary conditions. Taking the slab as a reference domain, we first studied the magnetization profiles at the upper critical dimensions $d=3$, $4$, $6$. We put the zeros and poles of the Weierstrass elliptic solutions (written as a function of a complex variable) in connection with the different boundary conditions.  We found general formulas in terms of Weierstrass elliptic functions, extending known results and finding new ones for percolation.

We then studied the connection between the critical magnetization profiles in a general domain and the solutions of the Yamabe problem in
the same domain and with the same boundary conditions.

The solutions of the Yamabe equation solve the so-called Yamabe problem, which amounts to finding a metric having constant curvature. This shows that the saddle-point equation for the critical magnetization at the upper critical dimension is equivalent to a purely geometric description, where the euclidean metric is conformally altered to obtain a space with uniform negative curvature. 
In the slab geometry, using the Yamabe equations we derived analytical expressions for critical magnetization profile as inverse hypergeometric functions valid for any dimension $d \ge 2$. The expressions are valid for fixed boundary conditions, but with suitable shifts in the argument of the slab coordinate variable we can obtain the corresponding solutions for the other considered boundary conditions. Lattice mean-field results for the slab geometry in $d=4$ have been also presented.

The magnetization profiles in the specific case of the four-dimensional Ising model with fixed boundary conditions in the slab are compared with Monte Carlo simulations, finding good agreement. 

The presented results explicitly confirm at the upper critical dimension recent results presented in \cite{gori}. There, general correlation functions of the magnetization field in a bounded domain have been conjectured to be related to the solution of a fractional Yamabe problem, which with vanishing anomalous dimension becomes the integer problem here considered. 

As next step at the upper critical dimension, a worthwhile subject of investigation would be to study spin-spin correlation functions in the four-dimensional case for bounded domains, and check if and how they depend on the geometric distance defined by the metric obtained from the solution of the integer Yamabe problem in the considered domain. 
{Another important question is how the present approach can treat or be extended to other surface universality classes, such as the special transitions \cite{Diehl1997}. We also observe that the present paper confirms that the Yamabe approach correctly describes fixed boundary conditions, while more work is needed to understand whether it can be extended to other boundary conditions such as open boundary conditions.}

{The well-known statistical models have a small anomalous dimension in three dimensions. Hence, one could be led to the study of a fractional Yamabe problem in which the exponent of the Laplacian is close to an integer value, to try and obtain a solution as a perturbation around the solution of the integer Yamabe equation.} %Next, one would like to study the case of small anomalous dimensions, which would lead to the study of a fractional Yamabe problem with the Laplacian entering with a fractional exponent, near to an integer value. 
This would be interesting both to compare with other perturbative approaches and for the challenging task to develop a suitable perturbative schemes to solve the fractional Yamabe equation relevant for critical models below the upper critical dimension. 
{Its effectiveness would then have to be compared with standard perturbative techniques.}
%{Whether this perturbative approach based on the fractional Laplacian can be concretely pursued and be similarly or more effective than the standard perturbative techniques remains to be seen.}

We also wonder if the analytic structure found can be extended
beyond the slab geometry at the upper critical dimension treated here. 
When the domain is not of the form $[0,L]\times \mathbb{R}^{d-1}$, is it still possible, given a magnetization profile, to obtain profiles for other boundary conditions by some process of continuation? 
%In a general domain, the magnetization depends on more than one coordinate: is there a way to obtain magnetization functions $m(\mathbf{x})$, for different boundary conditions, by some process of continuation of the one-dimensional profiles? \\ %When the domain is not morally a segment but a general shape can different boundary conditions be found by some process of continuation? 

If we instead consider the fractional Yamabe problem (in the slab to begin with) can we use similar techniques to the ones described here to provide a solution? The solution
to this problem was already found in~\cite{gori} via generalization of AdS/CFT-borne scattering techniques~\cite{Graham2003} providing a beautiful mathematical framework. When searching for explicit solutions to the fractional Yamabe problem within this framework, however, one is faced with great mathematical challenges, as it requires to both solve the vacuum Einstein equations for a metric in $d+1$ dimensions, and then find the solution of a nonlinear eigenvalue problem in the obtained metric space. It would be very appealing to have simpler schemes available.%We have shown how the description of bounded critical %phenomena through the introduction of a uniformizing metric is consistent with the mean field description at the upper critical dimension. From this, we obtained order parameter profiles in a slab for different boundary conditions.

More concretely, the next step will be to study a three dimensional system, a case which is also numerically more accessible, to test the predictions of the fractional Yamabe equation against data below $d_c$. It is then worth asking whether the profiles obtained from $\gamma(\mathbf{x})$ and the unknown function that describes two-point correlations can be approximately recovered in a perturbative manner, as an expansion around the four dimensional solutions.

\vspace{0.5cm}
{\it Acknowledgements:} We thank J. Cardy,
 N. Defenu, S. Dietrich, T. Enss, A. Gambassi and A. Squarcini for useful discussions at various stages of this work. GG is supported by the Deutsche Forschungsgemeinschaft (DFG, German Research Foundation) under Germany’s Excellence Strategy EXC 2181/1 -- 390900948 (the Heidelberg STRUCTURES Excellence Cluster).

\appendix

\section{Reminders about Weierstrass elliptic functions}
\label{app:Wei}

Elliptic functions appear in numerous areas of physics. They get their name from their property of being the inverse of elliptic integrals. A complex function of one complex variable $f(z)$ is called elliptic if it is meromorphic (its only singularities are poles) and is doubly periodic,

\begin{equation}
f(z+2\omega_1)=f(z), \qquad f(z+2\omega_2)=f(z),
\end{equation}

with $\omega_1/\omega_2 \notin \mathbb{R} $. $\omega_1$ and $\omega_2$ are %predictably 
called half-periods. The double periodicity induces a tessellation of the complex plane in parallelograms. It is therefore sufficient to know the values of the function within one of these parallelograms, say the one with vertices $0$, $2\omega_1$, $2\omega_2$ and $2\omega_1+2\omega_2$.

The Weierstrass $\wp$ function is probably the most intuitive elliptic function to construct, starting from the definition and the requirement of having only a double pole within each cell:
\begin{equation}
\begin{split}
    \wp (z)= \frac 1{z^2}+\sum_{\substack{m,n=-\infty\\ (m,n)\neq (0,0) } }^{\infty} \left( \frac 1{(z+2\,m\,\omega_1+2\,n\,\omega_2)^2}\right.- \\
    -\left.\frac 1{(2\,m\,\omega_1+2\,n\,\omega_2)^2}
    \right).\qquad
\end{split}
\end{equation}

Instead of the half periods, the Weierstrass function can be identified with another pair of numbers, $g_2$ and $g_3$, called invariants. They are the lowest order coefficients in the Laurent expansion of $\wp$ around $0$:

\begin{equation}
\wp (z) = \frac 1 {z^2}+\frac{g_2}{20} z^2 + \frac{g_3}{28}z^4 + O(z^6).
\end{equation}

The invariants can be obtained from the half periods as

\begin{equation}
\begin{split}
g_2=60\sum_{(m,n)\neq(0,0)}\frac 1{(2m\omega_1+2n\omega_2)^4} \\
g_3=140\sum_{(m,n)\neq(0,0)}\frac 1{(2m\omega_1+2n\omega_2)^6}.
\end{split}
\end{equation}

The importance of the invariants comes from the fact that they appear in the differential equation that the Weierstrass function obeys:

\begin{equation}
\wp'^2(z)=4\wp ^3(z)-g_2 \wp (z) - g_3=0.
\end{equation}

The particular cases encountered in the text are:
\begin{itemize}
    \item $(g_2,g_3)=(1,0)$, called lemniscatic elliptic function $\wp_l$, which gives orthogonal semiperiods $\omega_1=\omega_l=\Gamma(\frac 1 4)^2/(4\sqrt{\pi})$, $\omega_2=i \omega_1$, that tessellate the complex plane with squares.
\item $(g_2,g_3)=(0,1)$, the equiharmonic case $\wp_e$, with $\omega_1=\omega_e=\Gamma(\frac 1 3)^3/(4\pi)$ and $\omega_2=\frac 1 2(\sqrt 3 i-1)\omega_1$, which produces a tessellation of parallelograms each made of two equilateral triangles.
\end{itemize}
%\vspace{.4cm}

\section{Inverse functions for the lemniscatic and
equiharmonic Weierstrass elliptic functions}\label{appB}
%\printbibliography
In the main text explicit solutions
for the nonlinear ODE 
$m''(x)\propto m^\frac{d+2}{d-2}$
have been derived for special values of
$d$ in terms of the lemniscatic and equiharmonic Weierstrass
elliptic functions $\wp_l$ and $\wp_e$.
When solving the equation for generic $d$
a solution has been presented in implicit form.
This allows us to write the following 
inversion formulas for $\wp_l$ and $\wp_e$:
\begin{equation}
    {\wp_l}^{-1}(x)=\frac{1}{\sqrt{x}} {\,\,}_2F_1\left(\frac{1}{4},\frac{1}{2};\frac{5}{4};\frac{1}{4x^2}\right)
\end{equation}

\begin{equation}
    {\wp_e}^{-1}(x)=\frac{1}{\sqrt{x}} {\,\,}_2F_1\left(\frac{1}{6},\frac{1}{2};\frac{7}{6};\frac{1}{4x^3}\right).
\end{equation}
These two results can be obtained
by considering different representation
of the elliptic integrals $\int \mathrm{d}x \frac{1}{\sqrt{4x^3-1}}$
and  $\int \mathrm{d}x \frac{1}{\sqrt{4x^3-x}}$.

\section{Open boundary conditions with a scaling magnetic field}
\label{app00}
%\nocite{*}

In the main text we did not consider the possibility of leaving the spins on both boundaries to take any value, that is open boundary conditions (OBC, which we will label $00$). The reason is that in this case, the magnetization profile is trivially $m(x)=0$ throughout the slab since the system is at the critical temperature (see, e.g. \cite{Zandi2007}). In order to obtain nontrivial magnetization profiles for OBCs, one needs to introduce a magnetic field. The action at the critical point thus reads:

\begin{equation}
 S=\int d^dx\left[ \frac 1 2\phi(-\bigtriangleup)\phi  + 
 g\, c_d\, (\phi^2)^{\frac{d}{d-2}} -  h(x)\phi(x) \right],\label{actionh}\end{equation} 
and the corresponding saddle-point equation is

\begin{equation}\label{campoest}
  {  m''(x)=g \frac {d(d-2)}{4} m(x)^{\frac{d+2}{d-2} }-h(x).}
\end{equation}

\begin{figure}
\centering
    \includegraphics[scale=.65]{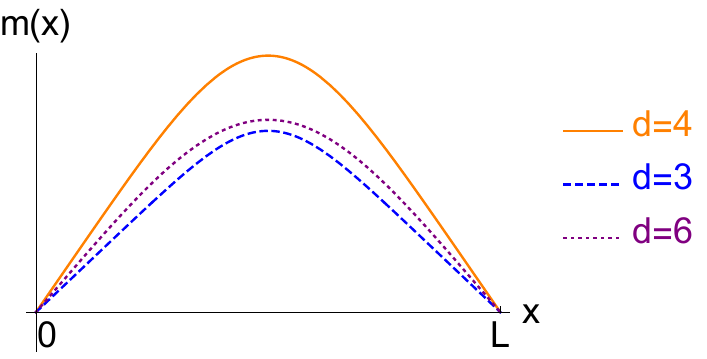}
    \caption{Magnetization profiles with OBC and a scaling magnetic field, in the three cases.}
    \label{profili00}
\end{figure}

The scaling dimension of the magnetic field is given by $\Delta_h=d-\Delta_{\phi}$. In the language of the Yamabe approach,  $\gamma(x)$ takes the role of the point-dependent length scale of the system, so a sensible choice would be to introduce a \textit{scaling} magnetic field, given by an appropriate power of the conformal factor: $m(x)\propto \gamma(x)^{\Delta_{\phi}-d}$. One can then use \eqref{reshape} to replace $\gamma(x)$ with the magnetization profile itself. Remembering that $\Delta_{\phi}=\frac{d-2}2$, we get
\begin{equation}
    h(x)\propto m^{\frac{d+2}{d-2}}.
\end{equation}
In particular, choosing
\begin{equation}\label{hm}
        h(x)=g\frac{d(d-2)}2  m^{\frac{d+2}{d-2}}
\end{equation}
we get that the solutions of \eqref{campoest} are also solutions of the saddle-point equation without external field, but with $g$ in \eqref{saddle_point} replaces by $-g$. This means that these solutions also take the form of elliptic functions: in particular, they are obtained joining two zeros in the complex plane, as shown in Fig. \ref{politutti}.

The magnetization profiles with the external field given by \eqref{hm} for $d=4,3,6$, plotted in Fig. \ref{profili00} are the following:
\begin{align}
        & d=4:    & m_{00}(x)=\wp_l(\omega_l + i\,x)^{1/2},   &\quad x\in[-\omega_l, \omega_l] &\nonumber
 \\
    & d=3:   & m_{00}(x)= \wp_e( x)^{-1/2},  & \quad x\in[0, 2\omega_e] &\nonumber \\
   & d=6:  & m_{00}(x)= \wp_e(\omega_e+i\,x) ,
& \quad x\in[-\frac{\omega_e}{\sqrt{3}},\frac{\omega_e}{\sqrt{3}}] &\nonumber
\end{align}

\onecolumngrid

%\begin{widetext}
\begin{figure}[t]
\begin{minipage}{.3\textwidth}
    \includegraphics[width=\textwidth]{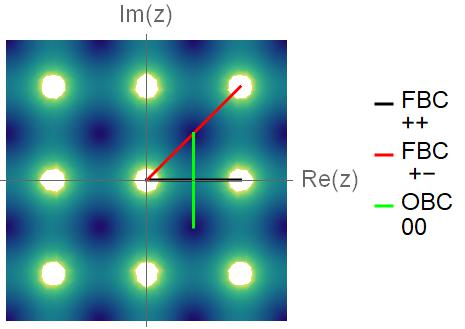}
\end{minipage} \qquad \begin{minipage}{.3\textwidth}
    \includegraphics[width=\textwidth]{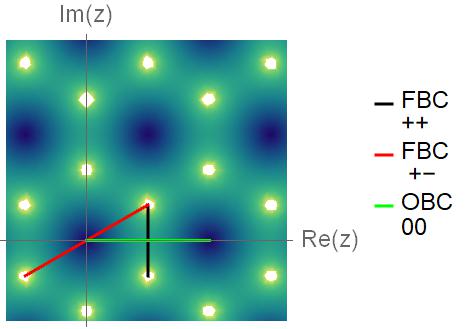}
\end{minipage} \qquad
\begin{minipage}{.3\textwidth}
    \includegraphics[width=\textwidth]{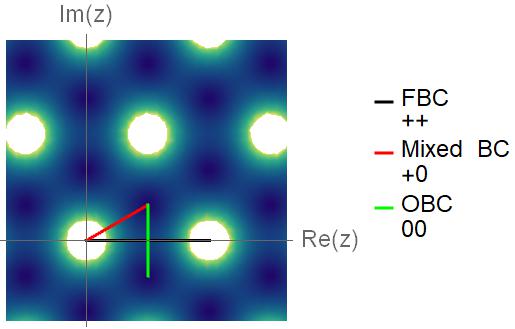}
\end{minipage}
\caption{Pole and zero structure of the solutions of the saddle-point equations in $d=4,3,6$ respectively, with the addition of the $00$ profile. The $d=3$ and $d=6$ structures are dual to each other: the $00$ solution in $d=3$ corresponds to the $++$ solution in $d=6$ and vice versa.}\label{politutti}
\end{figure}
\twocolumngrid

Other magnetic fields could be chosen, however the corresponding magnetization profile would in general no longer be an elliptic function. 

Let us focus on the $d=4$ solution. Numerical solutions on the Ising model with an external field confirmed the correctness of the solution $m_{00}$. The function $m_{00}$ is nothing but the reciprocal of the $++$ solution \eqref{bc}, up to a constant: shifting the argument of the function by half a period has the same effect as taking its inverse, since both operations swap poles and zeros. This is because the fundamental domain of $\wp_L$ is a square, and the square lattice tessellation of the plane is self-dual, as seen in the left panel of Fig. \ref{politutti}. 
This peculiarity of the elliptic functions must reflect in a property of the saddle-point equation, valid only for $d=4$ and when the Laplacian reduces to an ordinary second derivative: if $\phi(x)^{-1}$ is a solution, then so is ${\cal K}\, \phi(x)$, for an appropriate ${\cal K}$. Starting from
\begin{equation}
\frac {d^2}{dx^2} \phi(x)^{-1}=2 \phi(x)^{-3},
\end{equation}
we get
\begin{equation}
\phi \phi''-2(\phi')^2+2=0
\end{equation}
(where $\phi''=\frac {d^2}{dx^2} \phi$). 
Taking another derivative gives
\begin{equation}
\frac{\phi'''}{\phi''}=3 \frac{\phi'}{\phi},
\end{equation}
and, after integrating and exponentiating, we are left with
\begin{equation}
\phi''(x)={\cal K}\, \phi(x)^3,
\end{equation}
which is the saddle-point equation for $\sqrt{\frac 2 {\cal K}}\,\phi(x)$.%, with ${\cal K}$ an arbitrary constant.

Similarly, the $d=3$ and $d=6$ solutions are dual to each other, as can be seen in Fig. \ref{politutti}: the poles in $d=6$ correspond to the zeros in $d=3$ and vice versa, linking the $++$ solution in one case to the $00$ solution in the other.

Finally, we can discuss how to relate the 
result \eqref{gammad} to different boundary conditions in any dimension. We write the magnetization via \eqref{reshape} as $m_d(x)=\gamma_d(x)^{-\Delta_\phi}=\gamma_d(x)^{-\frac{d-2}{2}}$ where $\gamma_d(x)$ is given in \eqref{gammad}. If in $m_d(x)$ we 
set $d=3$, $4$, and $6$ and aptly
rescale the $x$ domain we recover the $++$
boundary magnetization profiles given
in Sec. \ref{magprofiles}. 
This can be achieved by using the
identities reported in Appendix~\ref{appB}. 
The other profiles, referring to different
boundary conditions, can be
found by taking $x$ in \eqref{gammad}
to be complex according to the following scheme:
\vspace{-.3cm}
\begin{align}
&m_{++}(x)= m_d(x)& x\in(0,2), \nonumber\\
&m_{+0}(x)= m_d(x\,e^{i\,\pi/d}) e^{-i\,\pi \Delta_\phi/d}
&x\in(0,\sec(\pi/d)], \nonumber \\
&m_{00}(x)= m_d(1+i\,x) 
&\hspace*{-1cm}x\in[-\tan(\pi/d),\tan(\pi/d)],\nonumber
\end{align}
\\
{where, again, $m_{00}(x)$ solves the saddle-point equation with opposite sign.}
It is worth observing that the
three BCs can be found by evaluating
$\gamma$ on the right triangle 
$\mathbf{T}=(0,1,1+i\tan(\pi/d))$. Indeed the function
$z(w)=w^{1/d}{}_2F_1\left( \frac 1d,\frac 12; 1+\frac 1d; w\right)$ appearing in \eqref{gammad} is nothing but the
Schwarz function mapping the upper-half-plane ($w$ variable)
onto $\mathbf{T}$ ($z$ variable, {called} $x$ when real) keeping $z=0,\,1$ fixed.
In the convenient variable {$w$} we have that
$\gamma=w^{1/d}$ and $m=w^{-\Delta_\phi/d}=w^{-\frac{d-2}{2d}}$. 
The above consideration should shed some light on the appearance of the Weierstrass functions  $\wp_e$, $\wp_l$ in Sec.~\ref{magprofiles} and on the peculiarity of $d=3,4,6$: for those values, copies of the triangle $\mathbf{T}$ create a regular tessellation of the complex plane.

\bibliography{bibGRAFFE}

\end{document}